\documentclass[12pt]{article}
\usepackage{graphicx,multicol}
\usepackage{amssymb,amsmath}
\usepackage{color}

\newcommand{\Lcal}{\mathcal{L}}
\newcommand{\Ncal}{\mathcal{N}}
\newcommand{\Ecal}{\mathcal{E}}

\usepackage[normalem]{ulem}

\begin{document}
\begin{titlepage}
\hfill
\vbox{
    \halign{#\hfil         \cr
           arXiv:0810.1661  \cr
           } 
      }  
\vspace*{20mm}
\begin{center}
{\Large \bf Comments on Holographic Fermi Surfaces}

\vspace*{15mm}
\vspace*{1mm}

{Hsien-Hang Shieh$^1$ and Greg van Anders$^{1,2}$}

\vspace*{1cm}

{${}^1$Department of Physics and Astronomy,
University of British Columbia\\
6224 Agricultural Road,
Vancouver, B.C., V6T 1Z1, Canada\\
${}^2$Michigan Center for Theoretical Physics, Randall Laboratory of Physics,\\
The University of Michigan, Ann Arbor, MI 48109-1040, USA}

\vspace*{1cm}
\end{center}

\begin{abstract}
Recently, a mechanism for the development of a fermi surface in a
holographic model of large $N$ QCD with a baryon chemical potential was
proposed. We examine similar constructions to determine when this mechanism
persists. We find a class of models in which it does.
\end{abstract}

\end{titlepage}
\section{Introduction} \label{intro}
The development of gauge/gravity duality
\cite{adscft1,adscft2,adscft3,adscftrev} has provided the opportunity to
study strongly coupled gauge theories. One of the most beautiful aspects of
this subject is that gauge theory phenomena take on a geometric character in
the dual gravity picture. For example, confinement was shown in
\cite{wittenconf} to be related to the Hawking-Page transition in gravity, and
chiral symmetry breaking was shown to be related to the geometry of brane
embeddings in \cite{adsqcd}.

Recently, there has been interest in studying holographic systems in backgrounds
with electromagnetic fields \cite{afjk1,ems,afjk2,jk,bll1},
and at finite baryon density \cite{mmmt,baryon,nakamura,matsuura}
and combinations thereof \cite{bll2}.

\begin{figure}
\begin{center}
\begin{picture}(0,0)%
\includegraphics{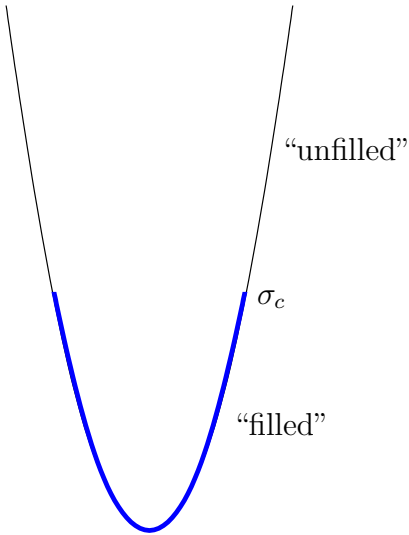}%
\end{picture}%
\setlength{\unitlength}{3947sp}%
\begingroup\makeatletter\ifx\SetFigFont\undefined%
\gdef\SetFigFont#1#2#3#4#5{%
  \reset@font\fontsize{#1}{#2pt}%
  \fontfamily{#3}\fontseries{#4}\fontshape{#5}%
  \selectfont}%
\fi\endgroup%
\begin{picture}(2490,3345)(-11,-2494)
\put(1726,-136){\makebox(0,0)[lb]{\smash{{\SetFigFont{12}{14.4}{\familydefault}{\mddefault}{\updefault}{\color[rgb]{0,0,0}``unfilled''}%
}}}}
\put(1426,-1861){\makebox(0,0)[lb]{\smash{{\SetFigFont{12}{14.4}{\familydefault}{\mddefault}{\updefault}{\color[rgb]{0,0,0}``filled''}%
}}}}
\put(1576,-1036){\makebox(0,0)[lb]{\smash{{\SetFigFont{12}{14.4}{\familydefault}{\mddefault}{\updefault}{\color[rgb]{0,0,0}$\sigma_c$}%
}}}}
\end{picture}%
\end{center}
\caption{This sketch of the mechanism shows string endpoints occupying some
region (shown in blue) below some critical point, labelled here by $\sigma_c$,
which is the location of the putative fermi surface. The region below
$\sigma_c$, in the interior of the space, is then considered the ``filled''
region and the ``unfilled'' region, toward the boundary of the space, is above.}
\label{sketch}
\end{figure}

In this note we further investigate a proposal made in \cite{baryon} for the
development of a fermi surface in a holographic model of large $N$ QCD. In that
paper a baryon chemical potential was added to the Sakai-Sugimoto
model \cite{adsqcd} by turning on the gauge field on the probe D8 brane.
This gauge field was sourced by string ends on the D8 brane and it was found
the minimum energy configuration with fixed baryon number had a sharp cutoff
in the positions of the string endpoints. As the baryon number was increased,
it was found that the position of the cut off moved from the interior toward
the boundary of the space. This is interesting because of a combination of two
factors: the position of the string ends are interpreted as fundamental quarks
in the holographic picture, and the interior of the space corresponds to the
infrared of the field theory whereas the boundary corresponds to the
ultraviolet. This suggests that the sharp cutoff in string endpoints is a
sharp cutoff in the energies of quarks, with increasing numbers of quarks
corresponding to an increase in the energy of the cutoff. This sort of behaviour
is what we would expect of a fermi surface for quarks, and the authors of
\cite{baryon} proposed that it is just that. See figure \ref{sketch} for a
sketch of this.

An interesting feature of this proposal is that in some sense the fermi
statistics were an emergent phenomena in the holographic description. They arose
from the mutual electrostatic repulsion of the string ends.

The system considered in \cite{baryon} was a $3+1$-dimensional gauge-theory
with massless fundamental quarks. In this note we investigate whether this
mechanism persists in other theories. There are two sources of motivation for
this. If this mechanism is peculiar feature of the Sakai-Sugimoto model, then it
should not be thought of as the formation of a fermi surface, but could be a
hallmark of some other interesting physics. One can, in principle, construct
other brane intersections that describe strongly coupled fermions. If these
systems do not exhibit this mechanism, that suggests that the mechanism is
not indicating the presence of a fermi surface. Conversely, if we can show that
this mechanism exists in a class of other models, then it is interesting to
understand what types of examples we can construct. Since strongly coupled
fermions are of interest in many areas of physics it is of interest to establish
that there are diverse examples of systems that exhibit this mechanism.

In section \ref{setup} we describe the set of systems we will consider, and
establish our notation and conventions. The systems include those arising from
brane intersections and $\Ncal=4$ SYM on $R\times S^3$. In section
\ref{dynamic} we will review the mechanism found in \cite{baryon}. We will then
look for this mechanism in a broad class of systems, and establish precise
conditions for when it does and does not occur. We will find that the conditions
for the non-formation of a putitive fermi surface at finite density are
extremely restrictive. The class of systems for which a putative fermi surface
does develop includes strongly coupled lower dimensional systems that may be
relevant for condensed matter physics. We discuss our findings
in section \ref{discussion}.

\section{Probe Brane Setup} \label{setup}
In order to explore this putative mechanism for a holographic fermi surface we
are interested in studying systems with fermions in the fundamental
representation of the gauge group. In this work we will restrict ourselves to
keeping the number of flavours fixed to one. To introduce this single flavour,
we will embed a single brane in the background geometry, and work in the probe
limit. In \cite{baryon}, the system under study was the Sakai-Sugimoto model
\cite{adsqcd} in which a probe D8-$\overline{\text{D8}}$ pair is embedded in
the background of $N_c$ D4 branes wrapped on a circle. Anti-periodic boundary
conditions are taken for the fermions around the circle, as proposed in
\cite{wittenconf}, so that, at energies much smaller than the Kaluza-Klein
scale, the adjoint sector is pure Yang-Mills theory. The resulting theory is
a $3+1$-dimensional gauge theory with fundamental fermions.

In this work, we are interested in understanding if the mechanism for a fermi
surface proposed in \cite{baryon} persists in other holographic constructions
of strongly coupled gauge theories. To this end we will consider two classes of
systems. The first class, including the Sakai-Sugimoto model, is constructed
from D(p+1)-Dq or D(p+1)-Dq-$\overline{\text{Dq}}$ configurations. In the second
class is $\Ncal=4$ SYM on $R\times S^3$ which is dual to global AdS.

\subsection{D(p+1)-Dq or D(p+1)-Dq-$\overline{\text{Dq}}$ Systems}
In this subsection we will discuss systems that can be constructed from
embedding a probe Dq brane in a D(p+1)-brane background. They are of interest
in the context of looking for systems that develop fermi surfaces because they
generically include light fermions that come from the string ground state in the
Ramond sector.
We will adopt conventions similar to those in
\cite{kmmw1,kmmw2,vw}. The metric and dilaton for these systems have the form
\begin{equation}
\begin{split}
ds^2 &=  \left(\frac{U}{R_p}\right)^\frac{6-p}{2}
	\left(\eta_{\mu\nu}dx^\mu dx^\nu+f(U)dx_{p+1}^2\right)
	+\left(\frac{R_p}{U}\right)^\frac{6-p}{2}
	\left(\frac{dU^2}{f(U)}+U^2 d\Omega_{7-p}^2\right),\\
e^\phi &= \frac{g^2_\text{YM}}{(2\pi)^{p-1}(\alpha')^\frac{p-2}{2}}
	\left(\frac{R_p}{U}\right)^\frac{(6-p)(2-p)}{4} \, ,
\end{split}
\end{equation}
where
\begin{equation}
\begin{split}
f(U) &= 1-\left(\frac{U_0}{U}\right)^{6-p} \, , \\
R_p^{6-p} &= g^2_\text{YM} N_c d_p (\alpha')^\frac{6-p}{2} \, , \\
d_p &= 2^{5-2p}\pi^\frac{6-3p}{2} \Gamma\left(\frac{6-p}{2}\right) \, .
\end{split}
\end{equation}
The quantity $U_0$ determines the size of the compact $x_{p+1}$ direction
according to
\begin{equation}
\frac{R_{KK}}{U_0} = \frac{4\pi}{6-p}\left(\frac{R_p}{U_0}\right)^\frac{6-p}{2}
\, .
\end{equation}
It is convenient to define a new coordinate $\rho$ according to
\begin{equation}
U^\frac{6-p}{2} = \rho^\frac{6-p}{2}+ \frac{U_0^{6-p}}{4\rho^\frac{6-p}{2}} \, ,
\end{equation}
so that the metric has the form
\begin{equation}
ds^2 =  \left(\frac{U}{R_p}\right)^\frac{6-p}{2}
	\left(\eta_{\mu\nu}dx^\mu dx^\nu+f(U)dx_{p+1}^2\right)
	+\left(\frac{R_p}{U}\right)^\frac{6-p}{2}\frac{U^2}{\rho^2}
	\left(d\rho^2+\rho^2 d\Omega_{7-p}^2\right).
\end{equation}
In this form it is clear that the transverse directions are conformal to flat
space, and to facilitate the brane embedding, we will write these coordinates
as
\begin{equation}
d\rho^2+\rho^2 d\Omega_{7-p}^2 = d\lambda^2 + \lambda^2 d\Omega_l^2
	+ dy^2 + y^2 d\Omega_{6-l-p}^2 \, ,
\end{equation}
where $\rho^2=\lambda^2+y^2$. In these coordinates we will take the brane to be
extended in the $\lambda$ and $\Omega_l$ directions and sit at a point on
$\Omega_{6-l-p}$.

We will embed a single Dq brane that intersects $m$ of the field theory space
directions, wraps $a=0,1$ of the Kaluza-Klein directions\footnote{If the
Dq brane does not wrap the Kaluza-Klein direction, we will assume that it
sits at a fixed point.}, as well as filling the $\lambda$ direction and an $S^l$
in the transverse directions.
If $a=0$ the brane configuration is of the form
\begin{equation}
\begin{tabular}{rccccccc}
  & $t$ & $p$  & & KK & $\lambda$  & & \\
\cline{3-4}
D(p+1) & $\times$ & $\times \cdots \times$ & $\times \cdots \times$ & $\times$ &
  $\bullet$ & $\bullet \cdots \bullet$ & $\bullet \cdots \bullet$ \\
Dq & $\times$ & $\times \cdots \times$ & $\bullet \cdots \bullet$ & $\bullet$ &
  $\times$ & $\times \cdots \times$ & $\bullet \cdots \bullet$ \\
$\overline{\text{Dq}}$ & $\times$ & $\times \cdots \times$ &
  $\bullet \cdots \bullet$ & $\bullet$ &
  $\times$ & $\times \cdots \times$ & $\bullet \cdots \bullet$ \\
\cline{3-3} \cline{7-7}
 & & $m$ & & & & $l$ & \\
\end{tabular}
\end{equation}
and if $a=1$ it is
\begin{equation}
\begin{tabular}{rccccccc}
  & $t$ & $p$  & & KK & $\lambda$  & & \\
\cline{3-4}
D(p+1) & $\times$ & $\times \cdots \times$ & $\times \cdots \times$ & $\times$ &
  $\bullet$ & $\bullet \cdots \bullet$ & $\bullet \cdots \bullet$ \\
Dq & $\times$ & $\times \cdots \times$ & $\bullet \cdots \bullet$ & $\times$ &
  $\times$ & $\times \cdots \times$ & $\bullet \cdots \bullet$ \\
\cline{3-3} \cline{7-7}
 & & $m$ & & & & $l$ & \\
\end{tabular}
\end{equation}
The Born-Infeld action for the embedded brane is
\begin{equation}
S=-\mu_q \omega_l R_{KK}^a \int d^{m+a+1}x \int d\lambda H(\rho) \lambda^l
	\sqrt{1+y'^2-\frac{\rho^2}{U^2}\tilde A'^2} \, ,
\end{equation}
where $\omega_l$ is the volume of $S^l$, $\tilde A=2\pi\alpha' A_0$, and
\begin{equation}
H(\rho) = \left(1-\left(\frac{U_0}{U}\right)^{6-p}\right)^\frac{a}{2}
	\left(\frac{U}{\rho}\right)^{l+1}
	\left(\frac{U}{R_p}\right)^\frac{(6-p)(4-\#\text{ND})}{4} \, ,
\end{equation}
where
\begin{equation}
\#\text{ND}=p+l-m-a+2 \,
\end{equation}
is the number of Neumann-Dirichlet directions. Systems with $a=0$ will be
D(p+1)-Dq-$\overline{\text{Dq}}$ systems and those with $a=1$ will be D(p+1)-Dq
systems.

To introduce baryons, we will use the idea of \cite{wittenbaryon}. In each
D(p+1) brane background is an $S^{7-p}$ carrying flux. We will wrap D(7-p)
branes on this sphere, and to satisfy the Gauss law constraint on the sphere
there must be $N_c$ strings ending on each D(7-p) brane. The other ends will
sit on the Dq brane and provide a source for electric flux on the brane. 

A D(7-p) brane at a position $U$ has the Born-Infeld action
\begin{equation}
S=-\mu_{7-p} R_p^{6-p} \omega_{7-p} \int dt U \, .
\end{equation}
The action for a collection of baryons with density $\rho_B(\lambda,x)$ is,
therefore,
\begin{equation}
S_B=-\mu_{7-p} R_p^{6-p} \omega_{7-p} \int d^{m+a+1}x
	\int d\lambda U\rho_B(\lambda,x) \, .
\end{equation}
In what follows we will take the baryon density to be homogeneous in the field
theory space directions, $\rho_B(\lambda,x)=\rho_B(\lambda)$. There is also the
effect of the string ends on the Dq brane, whereby the electromagnetic coupling
contributes a term to the action of the form
\begin{equation}
S_s=\frac{N_c}{2\pi\alpha'} \int d^{m+a+1} x
	\int d\lambda \rho_B(\lambda) \tilde{A} \, .
\end{equation}

\subsection{Global $AdS_5\times S^5$}
In this section we will consider D7 branes probing global $AdS_5\times S^5$.
We will use the Fefferman-Graham coordinates
\begin{equation}
ds^2 = R^2\left( -\frac14\left(\frac1z+z\right)^2 dt^2 + \frac{dz^2}{z^2}
	+\frac14\left(\frac1z-z\right)^2 d\Omega_3^2
	+d\theta^2 + \sin^2\theta d\phi^2 + \cos^2\theta d\bar\Omega_3^2 \right)
\end{equation}
where
$z=1$ is the centre of AdS, and the boundary is at $z=0$.

We will then consider the D7 embedding where it fills $AdS_5$ as well wrapping
an $S^3\subset S^5$. The DBI action for this embedding takes the form
\begin{equation}
S = -\Ncal \int d\Omega_3 \int d\bar\Omega_3 \int dt \int dz
	\cos^3\theta f_1(z)
	\sqrt{f_2(z)\left(\frac1{z^2}+\theta'^2\right)-
		\tilde{A}'(z)^2}
\end{equation}
where
\begin{equation}
f_1(z)=\left(\frac12\left(\frac1z-z\right)\right)^3 \, , \qquad
        f_2(z)=\left(\frac12\left(z+\frac1z\right)\right)^2 \, ,
\end{equation}
$\Ncal = R^8 \mu_7$, and $\tilde{A}=2\pi\alpha' A_0/R^2$.
Here $R^4=4\pi g_s N_c\alpha'^2$ and $\mu_7=((2\pi)^7g_s\alpha'^4)^{-1}$.

Again this background carries flux, in this case through $S^5$. We will
introduce baryons by wrapping D5 branes on this $S^5$. A D5 brane at position
$z$ will contribute to the action
\begin{equation}
S = -R^6\mu_5 \omega_5 \int dt \frac12\left(z+\frac1z\right)
\end{equation}
The action for a collection of baryons with density $\rho_B(z,t,\Omega_3)$ is,
therefore,
\begin{equation}
S_B = R^6\mu_5 \omega_5 \int dt \int d\Omega_3 \int dz
	\frac12\left(z+\frac1z\right)\rho_B(z,t,\Omega_3)
\end{equation}
Again, we will take configurations that are homogeneous in the field theory
directions so that $\rho_B(z,t,\Omega_3)=\rho_B(z)$. There is also the effect
of the string endpoints that has the form
\begin{equation}
S_s = \frac{N_c R^2}{2\pi\alpha'}
	\int dt \int d\Omega_3 \int dz \tilde{A}(z) \rho_B(z)
\end{equation}

\section{Finding the Fermi Surface} \label{dynamic}
\subsection{General Considerations} \label{gencon}
In this section we will investigate the brane constructions we set out in the
last section in the presence of baryons. We begin by noting that in all of the
brane constructions we presented above, the Born-Infeld action for the probe
brane took the form\footnote{We have absorbed the overall constant in $f$, as
it will not play a role in our analysis.}
\begin{equation} \label{sgen1}
S = -\int d\sigma f(y,y',\sigma)\sqrt{1-g(y,y',\sigma)A'^2} \, .
\end{equation}
There are additional contributions to the action from the string end points
and the masses of the branes
\begin{equation} \label{sgen2}
S' = N_c \int d\sigma \rho_B A - \int d\sigma \rho_B M(y,\sigma) \, .
\end{equation}
Altogether, these give an equation of motion for the gauge field as
\begin{equation}
\frac{d}{d\sigma} \frac{\partial \Lcal}{\partial A'} = N_c \rho_B \, .
\end{equation}
It will be convenient to use the electric flux
\begin{equation}
E \equiv \frac{\partial \Lcal}{\partial A'} =
	\frac{fgA'}{\sqrt{1-gA'^2}} \, ,
\end{equation}
or equivalently
\begin{equation}
gA'^2=\frac{E^2/g}{f^2+E^2/g} \, ,
\end{equation}
which means the equation of motion for $A$ is
\begin{equation} \label{Eeom}
E' = N_c \rho_B \, .
\end{equation}
There will be two generic cases we would like to consider, either the probe
brane extends from $\sigma=\infty$ to $\sigma=-\infty$, which we will call case
1, or ends at some $\sigma=\sigma_0$, which we will call case 2.\footnote{We
are to free reparametrize $\sigma$ so that $\sigma_0=0$ for convenience.} We can
integrate the equation of motion for $E$ to find that in case 1
\begin{equation}
2E_\infty = n_B N_c \, ,
\end{equation}
or in case 2
\begin{equation}
E_\infty = n_B N_c \, .
\end{equation}
Putting this together we find
\begin{equation}
S'=\int d\sigma E' A - \frac1{N_c} \int d\sigma E' M(y,\sigma) \, ,
\end{equation}
and integrating by parts gives
\begin{equation}
S'=2\mu E_\infty-\int d\sigma \frac{E^2/g}{\sqrt{f^2+E^2/g}}
	- \frac1{N_c} \int d\sigma E' M(y,\sigma) \, ,
\end{equation}
for case 1, or
\begin{equation}
S'=\mu E_\infty -\int d\sigma \frac{E^2/g}{\sqrt{f^2+E^2/g}}
	- \frac1{N_c} \int d\sigma E' M(y,\sigma) \, ,
\end{equation}
for case 2. Similarly, substituting $A'$ in terms of $E$ in $S$ gives
\begin{equation}
S = -\int d\sigma \frac{f^2}{\sqrt{f^2+E^2/g}} \, ,
\end{equation}
so that the total action is 
\begin{equation}
S_\text{total} = -\int d\sigma \sqrt{f^2+E^2/g}
	+2\mu E_\infty-\frac1{N_c} \int d\sigma E' M(y,\sigma) \, ,
\end{equation}
for case 1, or
\begin{equation}
S_\text{total} = -\int d\sigma \sqrt{f^2+E^2/g}
	+\mu E_\infty-\frac1{N_c} \int d\sigma E' M(y,\sigma) \, ,
\end{equation}
for case 2.
We have eliminated the gauge field entirely from the action by expressing it
in terms of the baryon charge density. We therefore find that the energy
density is either
\begin{equation}
\Ecal=\int d\sigma \sqrt{f^2+E^2/g}
	-2\mu E_\infty+\frac1{N_c} \int d\sigma E' M(y,\sigma) \, ,
\end{equation}
for case 1, or
\begin{equation}
\Ecal=\int d\sigma \sqrt{f^2+E^2/g}
	-\mu E_\infty+\frac1{N_c} \int d\sigma E' M(y,\sigma) \, ,
\end{equation}
for case 2.

The system will seek the minimum energy configuration, so we would like to
minimize this energy against the baryon charge density, subject to the
constraint that the overall baryon number is fixed. To ensure local
minimization we vary with respect to $E$ and find as a result that
\begin{equation} \label{Esoln}
\frac{E^2}g = \frac{f^2 g M'^2/N_c^2}{1-gM'^2/N_c^2} \, .
\end{equation}
It is this condition we will analyze to determine when a putative fermi surface
will develop.

\subsection{The Sakai-Sugimoto Model} \label{SS}
To understand how the putative fermi surface arises we will first
consider the simplest example, the Sakai-Sugimoto model at non-zero baryon
chemical potential. This has been considered previously in \cite{baryon}, in
which the mechanism was first proposed. It will be useful to reconsider what
happens in this simplest case, and to describe it in our notation before passing
on to more general considerations.

The Sakai-Sugimoto model is constructed by starting with D4 branes compactified
on a circle with anti-periodic (supersymmetry breaking) boundary conditions for
the fermions. At low energies the theory is then weakly coupled pure Yang-Mills
theory with a large number of colours $N_c$ \cite{wittenconf}. In the
supergravity limit it is described by the background geometry
\begin{equation}
\begin{split}
ds^2 &=  \left(\frac{U}{R_3}\right)^\frac{3}{2}
	\left(\eta_{\mu\nu}dx^\mu dx^\nu+f(U)dx_{4}^2\right)
	+\left(\frac{R_3}{U}\right)^\frac{3}{2}\frac{U^2}{\lambda^2}
	\left(d\lambda^2+\lambda^2 d\Omega_{4}^2\right),\\
e^\phi &= \frac{g^2_\text{YM}}{(2\pi)^{2}(\alpha')^\frac{1}{2}}
	\left(\frac{U}{R_3}\right)^\frac{3}{4} \, ,
\end{split}
\end{equation}
where\footnote{We've set $U_0=1$ for convenience.}
\begin{equation}
\begin{split}
f(U) &= 1-\frac{1}{U^3} \, , \\
U^\frac32 &= \lambda^\frac32+\frac{1}{4\lambda^\frac32} \, .
\end{split}
\end{equation}
This geometry is what one gets by putting $p=3$ in the general setup above.
Flavour is introduced by placing D8 and $\overline{\text{D8}}$ branes at the
antipodal points on the compact circle, $x_4$. In the limit that the number of
D8-$\overline{\text{D8}}$ pairs $N_f$ is much less than the number of colours,
$N_f\ll N_c$, the D8-$\overline{\text{D8}}$ can be treated as probing the
geometry of the D4 branes and backreaction can be neglected. The action for the
probe is given by
\begin{equation}
  S = -\mu_8 \omega_4 R_3^\frac32\int d^4x\int\frac{d\lambda}{\lambda}
       U^\frac72\sqrt{1-\frac{\lambda^2}{U^2}\tilde{A}'^2} \, .
\end{equation}
The field theory this setup describes at low energies, relative to the
compactification scale, is a 3+1-dimensional gauge theory with only gluons in
the adjoint sector and $N_f$ flavours of fermions in the fundamental
representation.

Baryons are introduced by wrapping D4 branes on the transverse $S^4$, and we
will consider a density of baryons in the field theory directions, giving
the action
\begin{equation}
  S_B = -\mu_4R_3^3\omega_4\int d^4x \int d\lambda U\rho_B(\lambda) \, .
\end{equation}
The final contribution comes from the interaction of the string ends with the
electromagnetic field on the D8 brane which gives the action
\begin{equation}
  S_s=\frac{N_c}{2\pi\alpha'}\int d^4x\int d\lambda\rho_B(\lambda)\tilde{A} \, .
\end{equation}
The various terms making up the total action are of the general form in
\eqref{sgen1} and \eqref{sgen2} of the previous section.

In section \ref{setup} we considered a general class of systems of probe branes
embedded in D-brane backgrounds. In the notation of section \ref{setup}, the
Sakai-Sugimoto model is a system with $p=3$, $q=8$, $a=0$, $m=3$, $l=4$, and
$\#\text{ND}=6$. In that notation then $\rho=\lambda$, and we can set $y=y'=0$
since it does not appear. This is because the D8 brane wraps
the entire transverse $S^4$. Because of this, the Sakai-Sugimoto model is the
simplest setting to look for this fermi surface. The form of the embedding, in
terms of the functions $f$, $g$ and $M'$ given in section \ref{gencon}, is then
\begin{equation}
  \begin{split}
    f(\lambda)&\propto \frac{U^{7/2}}{\lambda} \, , \\
    g(\lambda)&=\frac{\lambda^2}{U^2} \, , \\
    M(\lambda)&\propto U \, .
  \end{split}
\end{equation}
We would like to now consider the form of the electric field that minimizes the
energy. It is given by \eqref{Esoln} with $f$ and $g$ as above and
\begin{equation}
  M'\propto \frac{U}{\lambda}\frac{4\lambda^3-1}{4\lambda^3+1} \, .
\end{equation}

Let us consider now what happens when $\lambda$ is large. Asymptotically
$f\to\infty$, $g\to1$, and $M'\to1$. Explicitly, this means that the form of the
electric field that minimizes the energy density is
\begin{equation}
  E \propto \lambda^{5/2}
\end{equation}
for large $\lambda$, and clearly grows without bound toward the boundary. 
However, the asymptotic value of $E$ determines the baryon number density, which
we are keeping fixed, so it is not consistent for it to grow without bound.
In fact, since $E$ must not decrease, once the electric field reaches the value
$E_\infty$ that sets the baryon number density, the best we can do to minimize
the energy is to set the electric field equal to its asymptotic value. This
means that there is some critical $\lambda$ past which $E$ is constant. Since
in this region $E$ is constant, then by \eqref{Eeom} $\rho_B$ must vanish in
this region. However, large magnitudes for $\lambda$ correspond to large
energies, so that having vanishing baryon density above the critical $\lambda$
means that the quarks are all below some sharp energy cutoff. This energy cutoff
was proposed in \cite{baryon} as the development of a quark fermi surface.

\subsection{Other D(p+1)-Dq-$\overline{\text{Dq}}$ Systems} \label{Dqbar}
The mechanism which was proposed in \cite{baryon} for the formation of a Fermi
surface, and reproduced above, occurred in one example among the brane
constructions we have considered in \ref{setup}. We would like to now understand
if this mechanism can be generalized to other systems. The motivation for this
is twofold. If we fail to reproduce this mechanisms in other systems with
strongly coupled fermions, that would suggest that this mechanism
represents some other physics than the formation of a fermi surface. Conversely,
if this mechanism can be reproduced in other systems with strongly coupled
fermions, it would be suggestive that it might indeed indicate the formation of
a fermi surface. In that case, it is interesting to understand in what settings
it persists, and in particular to determine if there might be any that may be of
relevance to condensed matter physics.

The systems we will consider can be divided into three cases, those for which,
asymptotically, $y'\to0$, $y'\to \text{constant}$, or $y'$
diverges. The Sakai-Sugimoto model, as considered in \cite{baryon} and the
previous subsection, has $y'=0$. We will first consider systems in this class.
They have the asymptotic behaviour
\begin{equation}
  f \sim \sigma^{l+\frac{(6-p)(4-\#\text{ND})}{4}} \, ,
\qquad g \sim 1 \, , \qquad M' \sim 1 \, .
\end{equation}
This implies that for large $\sigma$, the electric field that minimizes the
energy is given by, according to \eqref{Esoln}, 
\begin{equation}
  E \sim \sigma^{l+\frac{(6-p)(4-\#\text{ND})}{4}} \, ,
\end{equation}
which diverges when $l+\tfrac{(6-p)(4-\#\text{ND})}{4}>0$. In appendix \ref{ND4}
we show that systems with $\#\text{ND}=4$ have $y\to\text{constant}$
asymptotically, and therefore $y'\to0$. Therefore, when these systems have
$l\ge1$ the local minimum for $E$ diverges for large $\sigma$, and the same
mechanism for the development of a fermi surface, which was found in
\cite{baryon} and reproduced above, occurs.

Suppose, alternatively, that $l+\tfrac{(6-p)(4-\#\text{ND})}{4}\le0$. In the
case that the inequality is strict, this would indicate that asymptotic value
of the electric field vanishes. Since this asymptotic value of the electric
field dictates that the baryon density must also vanish, then the system does
not contain any baryons, and therefore we would not expect a fermi surface to
form.

In the marginal case of $l+\tfrac{(6-p)(4-\#\text{ND})}{4}=0$ the electric field
would asymptote smoothly to some finite value $E_\infty$. This would suggest
that the charge density vanishes smoothly as we approach the boundary. Though we
expect that generically most of the quarks will still sit at lower energy scales
their density in this case should be given by a smooth distribution that
vanishes at high energies. In particular, this would indicate that a fermi
surface is not forming, even though we may expect one a priori.

As a result, in the case that $y'\to0$ asymptotically, we may only find that
a putative fermi surface does not form, when we expect one to, if
\begin{equation} \label{case1nosurf}
  l+\tfrac{(6-p)(4-\#\text{ND})}{4}=0 \, .
\end{equation}
This can be rewritten in a more illuminating way. In the notation we have
introduced above, the fermions coming from introducing some Dq brane are
confined to an $m$-dimensional defect in a $p+1$-dimensional gauge theory.
Expressing \eqref{case1nosurf} in terms of these parameters, only
when the dimension of the defect is given by
\begin{equation}
  m=\frac{(p-2)(7-p-q)}{2(4-p)}
\end{equation}
can we not have a putative fermi surface at finite density when $y'\to0$
asymptotically. It is straightforward to check that in the examples we are most
interested in, $p=1,2,3$, that this is only possible if $m=0$. To emphasize, in
this case, only when the fermions are localized on a point-like defect is it
possible for the putative fermi surface to not develop.

Some remarks are in order. As we discuss in appendix \ref{ND4}, the asymptotic
separation of the branes sets the quark mass. The above analysis showing that
the putative mechanism for the fermi surface occurs was independent of the
asymptotic value of the brane separation, as long as it was finite. This
means that the mechanism is insensitive to the quark mass. In the system
considered in \cite{baryon} the quarks were massless, so we have shown this
mechanism is also viable for massive quarks. We also point out
that the systems for which this mechanism occurs include the $1+1$-dimensional
D2-D4-$\overline{\text{D4}}$ system, the $2+1$-dimensional
D3-D5-$\overline{\text{D5}}$ system, and the $3+1$-dimensional
D4-D6-$\overline{\text{D6}}$ system. These lower dimensional systems are
interesting because they may serve as useful toy models for condensed matter
physics.

Next, consider the marginal case in which $y'\to\text{constant}$ asymptotically.
The behaviour in this case is similar to that in the previous one; the
asymptotic behaviour of the functions $f$ and $g$ just gain constant
coefficients that depend on the asymptotic value of $y'$, but they scale with
$\sigma$ in the same way. This indicates that, again, the only way that we
could have the non-formation of a putative fermi surface in interesting systems
is if the fermions are confined to a point-like defect.

Finally, consider the other case that $y'$ diverges for large $\sigma$. Suppose
that, for large $\sigma$, $y'\sim\sigma^k$ for some $k>1$.\footnote{In general
we might also want to have some logarithmic dependence on $\sigma$ as well, but
this does not have any effect on our conclusions.} Asymptotically
we have the behaviour
\begin{equation}
  f \sim \sigma^{l+\frac{(6-p)(4-\#\text{ND})}{4}-k} \, ,
\qquad g \sim \frac{1}{\sigma^{2k}} \, , \qquad M' \sim 1 \, .
\end{equation}
This implies that the form of the electric field that minimizes the energy is,
according to \eqref{Esoln},
\begin{equation}
  E \sim \sigma^{l+\frac{(6-p)(4-\#\text{ND})}{4}-3k} \, ,
\end{equation}
which diverges when $l+\tfrac{(6-p)(4-\#\text{ND})}{4}>3k$. Systems that satisfy
this requirement will also exhibit the mechanism in \cite{baryon} for the
development of a putative fermi surface.

If we consider the in which $l+\tfrac{(6-p)(4-\#\text{ND})}{4}<3k$ then the
asymptotic electric field and consequently baryon density both vanish. This
happened in the previous two cases as well, and it is not surprising that a
fermi surface would not form under these circumstances. Only the marginal case
of $l+\tfrac{(6-p)(4-\#\text{ND})}{4}=3k$ is therefore interesting. Again,
writing this in terms of the dimensionality of the defect, the gauge theory and
the probe branes we find a putative fermi surface doesn't form only when the
degree of divergence is
\begin{equation}
  k=\frac{1}{12}\left(2m(4-p)-(p-2)(7-p-q)\right) \, .
\end{equation}
In this case we would still expect that most of the quarks would sit at lower
energy scales, but the distribution must vanish smoothly at higher energies.

Let us now summarize the main results of this subsection. We first considered
two cases in which the embedding function $y$ had either the asymptotic
behaviour $y'\to 0$ or $y'\to\text{constant}$. We found in these cases that the
only circumstance in which we would not have the formation of a fermi surface
at finite density is in a $p+1$-dimensional field theory with fermions coming
from a Dq flavour brane that are localized on an $m$-dimensional defect where
\begin{equation} \label{cond1}
  m=\frac{(p-2)(7-p-q)}{2(4-p)} \, .
\end{equation}
We also pointed out that for systems of interest, where $p\le 3$, can only be
satisfied by $m=0$, i.e.\ by fermions localized at a point.
We considered a further class of systems in which the embedding function had
the asymptotic behaviour $y'\to\infty$. We found that, if the asymptotic
behaviour was such that $y'\sim\sigma^k$ for some $k>1$ in a holographic
description of a $p+1$-dimensional field theory with fermions coming from a Dq
flavour brane that that are localized on an $m$-dimensional defect, only when
\begin{equation} \label{cond2}
  k=\frac{1}{12}\left(2m(4-p)-(p-2)(7-p-q)\right) \, ,
\end{equation}
would a fermi surface not form at finite density. The two conditions
\eqref{cond1} and \eqref{cond2} on the non-formation of a putative fermi
surface at finite density are quite restrictive and indicate that generically
we should expect one to form.

\subsection{D(p+1)-Dq Systems} \label{Dq}
The analysis in the previous section carries over almost unchanged to the case
of D(p+1)-Dq systems. Note that the only modification is that the function
$H(\rho)$ picks up a factor of
\begin{equation}
\sqrt{1-\left(\frac{U_0}{U}\right)^{6-p}} \, ,
\end{equation}
as does, therefore, the function $f$ of section \ref{gencon}. This factor
asymptotes to unity, so that the conditions \eqref{cond1} and \eqref{cond2}
found above hold for D(p+1)-Dq systems as well.

An interesting example system for this case is the D4-D6 system where the
D6 brane intersects two of the field theory directions. This system has a
$3+1$-dimensional gauge field with fundamental particles localized on a
$2+1$-dimensional defect.

\subsection{Global $AdS_5\times S^5$}
We would like to consider the asymptotics of this system. They have been
analyzed previously in \cite{ko1}. There it was shown that
\begin{equation}
\theta(z) \sim \theta_0 z + \theta_2 z^3 + 2\theta_0 z^3 \ln z + \cdots \, .
\end{equation}
These asymptotics are not altered by an asymptotically constant $E$, which we
demand to have fixed baryon number.\footnote{It can be checked that the
asymptotic value of $E$ only enters this expansion at order $z^9\ln z$.}
We need to also consider the asymptotics of the functions that appear in
\eqref{Esoln}, they are
\begin{equation}
M' \sim -\frac1{2z^2} \, , \qquad
g \sim 4 z^4 \, , \qquad
f \sim \frac{1}{16 z^5} \, .
\end{equation}
Together, these imply that $E$ diverges as $z\to0$. As before, this indicates
the development of a putative fermi surface. Note the dual field theory in this
case is $\Ncal=4$ SYM on $R\times S^3$, so this mechanism also works in the case
of a field theory on a compact space.

\section{Discussion} \label{discussion}
In this paper we have considered a broad class of holographic systems at finite
baryon chemical potential. This was motivated by a proposal in \cite{baryon} of
a mechanism for the development of a fermi surface in a holographic model of
large $N$ QCD. We found that the mechanism for the formation of a putative
fermi surface persists across a broad class of models. This suggests
that the mechanism is not a peculiar feature of the Sakai-Sugimoto model that
results from one of its particular features, e.g.\ massless quarks, it's
dimensionality, the fact it's fermions are not localized on a defect, or that
the embedding is simple because the flavour branes fill the transverse $S^4$.
The systems we have studied are therefore useful for two reasons. They give some
reason to believe that this mechanism may indeed describe the formation of a
fermi surface, because it does occur in a wide class of systems with strongly
coupled fermions at finite density. They are also interesting because they
comprise a variety systems including lower dimensional ones that could find
application in condensed matter physics for understanding strongly coupled
phenomena in which fermi surfaces play an important role.

If the mechanism we have described in this note does indeed indicate the
development of a fermi surface, the picture we have presented here is
somewhat rudimentary. The theory of strongly coupled fermi liquids is a well
developed subject (see, e.g.\ \cite{LLv9}), and there are general expectations
for what features systems such as those under study here should exhibit. One
such feature is the phenomenon of zero sound. Investigations of zero sound in
holographic settings at finite density have been carried out recently in
\cite{zerosound,kp1,kp2}. The zero sound modes in these cases were identified
with fluctuations of the probe flavour branes, and the investigations were
carried out without explicitly including a source for the baryon charge density.
A fermi surface has not been identified in the absence of the sources, so this
might suggest that an explicit fermi surface is not important in finding zero
sound in holographic constructions. This seems confusing because zero sound is
associated with deforming the fermi surface. Another way of identifying a fermi
surface is by finding a pole in the retarded current-current correlation
function. Such an investigation was carried out for the Sakai-Sugimoto model in
\cite{kp2} but that analysis failed to reveal the expected pole. However, the
analysis was carried out in the absence of the putative fermi surface we
considered here, and it does not seem unreasonable that if a similar analysis
was carried out with it included, a sharp cutoff in the charge density might
produce such a pole.

Having now established the existence of a putative fermi surface in a diverse
set of examples, it would be particularly interesting to investigate what effect
the presence of the putative fermi surface has on the thermodynamics and the
spectrum of low energy excitations, as well as to look for a zero sound mode.
It is difficult, in the general framework we have used in this paper, to address
general features of the low energy spectrum for the whole class of systems under
consideration, for example the existence of poles in the retarded
current-current two-point function. It would interesting to revisit this
question in the Sakai-Sugimoto model as was done in \cite{kp2} to
determine if the existence of the putative fermi surface we have described above
is sufficient to produce the expected pole. Such a study is also of particular
interest in some of the lower dimensional examples that we have considered
because of their potential applications for condensed matter physics. A further
interesting question is to understand how the systems respond to external
electric and magnetic fields. We leave these investigations to future work.

\section*{Acknowledgements}
We would like to thank Borun Chowdhury, Samir Mathur, Jeremy Michelson,
Andy O'Bannon, and Gerardo Ortiz for interesting discussions, and to
Leo Pando Zayas for comments on the manuscript. We'd also especially like to
thank Mark Van Raamsdonk for many helpful discussions and comments on the
manuscript, and Jackson Wu for collaboration on related work. GvA would like to
thank the Ohio State University and the University of Chicago for hospitality
when this work was initiated and the support of a Theodore E.\ Arnold Graduate
Fellowship from the University of British Columbia.

\appendix
\section{Systems with $\#\text{ND}=4$}\label{ND4}
In this section we will consider a set of supersymmetric systems. The systems
we will consider have $a=0$ and $\#\text{ND}=4$.

First consider the situation in which the gauge field on the Dp brane is not
turned on. This will allow us to determine the relationship between the
asymptotics of the embedding function and the masses of the quarks and the value
of the quark condensate in the gauge theory.

In this case it is convenient to make all the quantities in the action
dimensionless by taking $U_0=1$. We would like to consider embeddings in which
$y\to y_\infty$ as we take $\lambda\to\infty$. The equation of motion has the
asymptotic form
\begin{equation} \label{ND4:aeom}
\partial_\lambda (\lambda^l y')=
	\frac{l+1}{2}\lambda^{p+l-8} y \, .
\end{equation}
This equation can be recast as a Bessel equation, and has the general solution
\begin{equation}
y(\lambda)= A\frac1{\lambda^{\frac{l-1}{2}}}
	J_{\frac{l-1}{6-p}}
	\left(\frac{\sqrt{2(l+1)}}{6-p}\frac1{\lambda^{\frac{6-p}2}}\right)
	+B\frac1{\lambda^{\frac{l-1}{2}}}
	J_{-\frac{l-1}{6-p}}
	\left(\frac{\sqrt{2(l+1)}}{6-p}\frac1{\lambda^{\frac{6-p}2}}\right)
 \, ,
\end{equation}
where $J$ are the usual Bessel functions. To determine the asymptotic behaviour
we note that $J_\alpha(x)\sim x^\alpha$ for small $x$, which means that, for
$l>1$,\footnote{For $l\le1$ one of the solutions diverges for large $\lambda$
indicating that the equation \eqref{ND4:aeom} is only valid for $l>1$.}
term multiplied by $A$ goes to a constant for $\lambda \gg 1$, and the term
multiplied by $B$ goes like $\lambda^{-(l-1)}$.

Following the analysis in \cite{kmmw2} if we have the asymptotic behaviour
\begin{equation}
y(\lambda)\sim y_\infty + \frac{c}{\lambda^{l-1}}
\end{equation}
then the quark mass will be
\begin{equation}
m_q = \frac{U_0 y_\infty}{2\pi\alpha'}
\end{equation}
and the condensate will be
\begin{equation}
\left< \bar\psi \psi \right> =
	 -2\pi\alpha' \mu_q (l-1)\omega_l U_0^l c \, .
\end{equation}

If we now turn on the gauge field, when $l>1$, similar considerations give
\begin{equation}
\tilde A \sim \tilde\mu - \frac{\tilde c}{\lambda^{l-1}} \, ,
\end{equation}
where $\tilde\mu$ is related to the chemical potential $\mu$ by
\begin{equation}
\tilde\mu = \frac{1}{2\pi\alpha'}\mu \, ,
\end{equation}
and $\tilde c$ determines the number of baryons according to
\begin{equation}
\tilde{c} = \frac{n_B N_c}{2(l-1)\mu_q \omega_l} \, .
\end{equation}

\providecommand{\href}[2]{#2}\begingroup\raggedright\endgroup

\begin{thebibliography}{10}

\bibitem{adscft1}
J.~M. Maldacena, {\it The large {N} limit of superconformal field theories and
  supergravity},  {\em Adv. Theor. Math. Phys.} {\bf 2} (1998) 231--252,
  [\href{http://xxx.lanl.gov/abs/hep-th/9711200}{{\tt hep-th/9711200}}].

\bibitem{adscft2}
S.~S. Gubser, I.~R. Klebanov, and A.~M. Polyakov, {\it Gauge theory correlators
  from non-critical string theory},  {\em Phys. Lett.} {\bf B428} (1998)
  105--114, [\href{http://xxx.lanl.gov/abs/hep-th/9802109}{{\tt
  hep-th/9802109}}].

\bibitem{adscft3}
E.~Witten, {\it Anti-de {S}itter space and holography},  {\em Adv. Theor. Math.
  Phys.} {\bf 2} (1998) 253--291,
  [\href{http://xxx.lanl.gov/abs/hep-th/9802150}{{\tt hep-th/9802150}}].

\bibitem{adscftrev}
O.~Aharony, S.~S. Gubser, J.~M. Maldacena, H.~Ooguri, and Y.~Oz, {\it Large {N}
  field theories, string theory and gravity},  {\em Phys. Rept.} {\bf 323}
  (2000) 183--386, [\href{http://xxx.lanl.gov/abs/hep-th/9905111}{{\tt
  hep-th/9905111}}].

\bibitem{wittenconf}
E.~Witten, {\it Anti-de {S}itter space, thermal phase transition, and
  confinement in gauge theories},  {\em Adv. Theor. Math. Phys.} {\bf 2} (1998)
  505--532, [\href{http://xxx.lanl.gov/abs/hep-th/9803131}{{\tt
  hep-th/9803131}}].

\bibitem{adsqcd}
T.~Sakai and S.~Sugimoto, {\it Low energy hadron physics in holographic {QCD}},
   {\em Prog. Theor. Phys.} {\bf 113} (2005) 843--882,
  [\href{http://xxx.lanl.gov/abs/hep-th/0412141}{{\tt hep-th/0412141}}].

\bibitem{afjk1}
T.~Albash, V.~G. Filev, C.~V. Johnson, and A.~Kundu, {\it {Finite Temperature
  Large N Gauge Theory with Quarks in an External Magnetic Field}},  {\em JHEP}
  {\bf 07} (2008) 080, [\href{http://xxx.lanl.gov/abs/0709.1547}{{\tt
  arXiv:0709.1547}}].

\bibitem{ems}
J.~Erdmenger, R.~Meyer, and J.~P. Shock, {\it {AdS/CFT with Flavour in Electric
  and Magnetic Kalb-Ramond Fields}},  {\em JHEP} {\bf 12} (2007) 091,
  [\href{http://xxx.lanl.gov/abs/0709.1551}{{\tt arXiv:0709.1551}}].

\bibitem{afjk2}
T.~Albash, V.~G. Filev, C.~V. Johnson, and A.~Kundu, {\it {Quarks in an
  External Electric Field in Finite Temperature Large N Gauge Theory}},  {\em
  JHEP} {\bf 08} (2008) 092, [\href{http://xxx.lanl.gov/abs/0709.1554}{{\tt
  arXiv:0709.1554}}].

\bibitem{jk}
C.~V. Johnson and A.~Kundu, {\it {External Fields and Chiral Symmetry Breaking
  in the Sakai- Sugimoto Model}},  {\em JHEP} {\bf 12} (2008) 053,
  [\href{http://xxx.lanl.gov/abs/0803.0038}{{\tt arXiv:0803.0038}}].

\bibitem{bll1}
O.~Bergman, G.~Lifschytz, and M.~Lippert, {\it {Response of Holographic QCD to
  Electric and Magnetic Fields}},  {\em JHEP} {\bf 05} (2008) 007,
  [\href{http://xxx.lanl.gov/abs/0802.3720}{{\tt arXiv:0802.3720}}].

\bibitem{mmmt}
D.~Mateos, S.~Matsuura, R.~C. Myers, and R.~M. Thomson, {\it {Holographic phase
  transitions at finite chemical potential}},  {\em JHEP} {\bf 11} (2007) 085,
  [\href{http://xxx.lanl.gov/abs/0709.1225}{{\tt arXiv:0709.1225}}].

\bibitem{baryon}
M.~Rozali, H.-H. Shieh, M.~Van~Raamsdonk, and J.~Wu, {\it {Cold Nuclear Matter
  In Holographic QCD}},  {\em JHEP} {\bf 01} (2008) 053,
  [\href{http://xxx.lanl.gov/abs/0708.1322}{{\tt arXiv:0708.1322}}].

\bibitem{nakamura}
S.~Nakamura, {\it {Comments on Chemical Potentials in AdS/CFT}},  {\em Prog.
  Theor. Phys.} {\bf 119} (2008) 839--847,
  [\href{http://xxx.lanl.gov/abs/0711.1601}{{\tt arXiv:0711.1601}}].

\bibitem{matsuura}
S.~Matsuura, {\it {On holographic phase transitions at finite chemical
  potential}},  {\em JHEP} {\bf 11} (2007) 098,
  [\href{http://xxx.lanl.gov/abs/0711.0407}{{\tt arXiv:0711.0407}}].

\bibitem{bll2}
O.~Bergman, G.~Lifschytz, and M.~Lippert, {\it {Magnetic properties of dense
  holographic QCD}},  \href{http://xxx.lanl.gov/abs/0806.0366}{{\tt
  arXiv:0806.0366}}.

\bibitem{kmmw1}
M.~Kruczenski, D.~Mateos, R.~C. Myers, and D.~J. Winters, {\it Meson
  spectroscopy in {AdS}/{CFT} with flavour},  {\em JHEP} {\bf 07} (2003) 049,
  [\href{http://xxx.lanl.gov/abs/hep-th/0304032}{{\tt hep-th/0304032}}].

\bibitem{kmmw2}
M.~Kruczenski, D.~Mateos, R.~C. Myers, and D.~J. Winters, {\it Towards a
  holographic dual of large-${N}_c$ {QCD}},  {\em JHEP} {\bf 05} (2004) 041,
  [\href{http://xxx.lanl.gov/abs/hep-th/0311270}{{\tt hep-th/0311270}}].

\bibitem{vw}
M.~Van~Raamsdonk and K.~Whyte, ``To appear.''.

\bibitem{wittenbaryon}
E.~Witten, {\it {Baryons and branes in anti de Sitter space}},  {\em JHEP} {\bf
  07} (1998) 006, [\href{http://xxx.lanl.gov/abs/hep-th/9805112}{{\tt
  hep-th/9805112}}].

\bibitem{ko1}
A.~Karch and A.~O'Bannon, {\it {Chiral transition of N = 4 super Yang-Mills
  with flavor on a 3-sphere}},  {\em Phys. Rev.} {\bf D74} (2006) 085033,
  [\href{http://xxx.lanl.gov/abs/hep-th/0605120}{{\tt hep-th/0605120}}].

\bibitem{LLv9}
E.~M. Lifshitz and L.~P. Pitaevskii, {\em Statistical Physics, Part 2}.
\newblock Butterworth-Heinemann, Oxford, 1980.

\bibitem{zerosound}
A.~Karch, D.~T. Son, and A.~O. Starinets, {\it {Zero Sound from Holography}},
  {\em Phys. Rev. Lett.} {\bf 102} (2009) 051602,
  [\href{http://xxx.lanl.gov/abs/0806.3796}{{\tt arXiv:0806.3796}}].

\bibitem{kp1}
M.~Kulaxizi and A.~Parnachev, {\it {Comments on Fermi Liquid from Holography}},
   {\em Phys. Rev.} {\bf D78} (2008) 086004,
  [\href{http://xxx.lanl.gov/abs/0808.3953}{{\tt arXiv:0808.3953}}].

\bibitem{kp2}
M.~Kulaxizi and A.~Parnachev, {\it {Holographic Responses of Fermion Matter}},
  \href{http://xxx.lanl.gov/abs/0811.2262}{{\tt arXiv:0811.2262}}.

\end{thebibliography}
\end{document}